\documentclass[12pt]{article}

\usepackage[a4paper,margin=1in]{geometry}
\usepackage{amsmath,amssymb}
\usepackage{bm}
\usepackage{graphicx}
\graphicspath{{./}}
\usepackage{natbib}
\usepackage{setspace}
\usepackage{hyperref}
\usepackage{float}

\hypersetup{
  colorlinks=true,
  linkcolor=blue,
  citecolor=blue,
  urlcolor=blue
}

\title{%
From Columns to Heaps: Dimensionless Similarity with PSD-Distributed Damköhler Numbers and Dual-Porosity Flow
}

\author{%
Juan J. Segura\thanks{Universidad Andres Bello, Chile, juan.segura.f@unab.cl}
}

\date{\today}

\begin{document}

\maketitle

\begin{abstract}
This work develops a unified, dimensionless framework for comparing geometrically similar reacting porous-flow systems across scale, with emphasis on hydrometallurgical heap leaching, when particle size distribution (PSD) and intraparticle pore structure differ. Under dynamic similarity, the dimensionless liquid residence-time distribution (RTD) is identical, but differences in PSD and internal porosity break microscopic similarity. Using the shrinking-core model (SCM), the analysis shows how a PSD in particle diameter maps to a distribution of particle-scale Damköhler numbers that governs heap-averaged conversion.

Explicit PSD to Damköhler transformations are derived for (i) external film control, (ii) intraparticle diffusion control, and (iii) mixed control via additive rates. Dual-porosity hydrology relevant to sedimentary or strongly stratified ores is then incorporated by coupling SCM kinetics to mobile and immobile liquid domains, introducing additional dimensionless groups that describe interporosity exchange. Two numerical examples map a lognormal PSD into film- and diffusion-controlled Damköhler distributions and compare column/heap conversion for different PSDs. A practical workflow is outlined for hydrometallurgical column-test interpretation, combining tracer RTD calibration with PSD-aware kinetic fitting in dimensionless time. The framework clarifies why diffusion-controlled leaching is far more sensitive to PSD tails and dual-porosity structure than film-controlled leaching, and it identifies the compact set of dimensionless groups that must be matched to ensure similarity between laboratory columns and industrial heaps.
\end{abstract}

\section{Introduction}

Heap leaching is a key unit operation in the hydrometallurgical processing of 
copper, gold, uranium and other commodities, and is likely to remain so as 
lower-grade and more complex ores are brought into production 
\citep{DixonPetersen2003,DixonPetersen2004,HeapScaleup}. Industrial heaps are 
large, highly heterogeneous systems in which chemistry, flow and transport 
interact over a wide range of time and length scales. Despite the advances of 
the last two decades in column testing, 1D and 2D modelling, and field 
monitoring, scale-up from laboratory columns and on-off pads to multi-lift 
commercial heaps still relies heavily on engineering judgement and empirical 
factors \citep{HeapHydroDixon,HeapSim2D,Robertson2017,Miao2017,Blackmore2018}.

Two aspects are particularly important in practice. First, the hydrodynamics of 
drip-irrigated heaps---including unsaturated flow, channeling, and dual-porosity 
behaviour---control the residence time distribution (RTD) of the leaching 
solution, the extent of ponding and saturation, and the degree to which the 
solution actually contacts reactive mineral surfaces \citep{HeapHydroDixon,
PreferentialFlow2016,RTDDrip2023,Robertson2017,Miao2017,Miao2021}. Second, the 
mineral itself is distributed among a range of particle sizes and internal pore 
structures. The particle size distribution (PSD), internal porosity and the 
presence of product layers within particles jointly determine which 
rate-controlling mechanisms (external mass transfer, intraparticle diffusion, 
or surface reaction) are most important, and how much of the mineral becomes 
effectively ``locked'' in diffusion-limited domains 
\citep{Szekely1975,Fuerstenau2005,FoglerSCM,MorenoPulido2025}.

Most mechanistic heap-leach models developed for hydrometallurgical systems 
incorporate some combination of advection--dispersion flow, dual-porosity or 
dual-permeability structure, and shrinking-core or related intraparticle 
kinetics \citep{DixonPetersen2003,DixonPetersen2004,HeapScaleup,
Robertson2017,Miao2017,Miao2021,HeapSim2D}. These models are powerful, but 
their complexity can make it difficult to see clearly which dimensionless 
groups control behaviour, how particle-scale heterogeneities map to heap-scale 
responses, and what conditions must be matched between column tests and 
industrial heaps to ensure meaningful similarity.

The purpose of this work is to provide a compact, dimensionless framework to 
clarify these issues, with a focus on industrially relevant heap-leach systems. 
We proceed in three main steps:
\begin{enumerate}
  \item We discuss geometric and dynamic similarity at the reactor or heap 
        scale, and show how matching a small set of macroscopic dimensionless 
        groups (e.g.\ Peclet number, capillary and Bond numbers) leads to 
        similar dimensionless RTDs for the liquid phase in geometrically 
        similar heaps.

  \item Using the shrinking-core model (SCM), we show how a PSD induces a 
        distribution of particle-scale Damk\"ohler numbers under three 
        commonly encountered regimes: external film (convection) control, 
        intraparticle diffusion control, and mixed control. We derive explicit 
        expressions that map a given PSD to a distribution of Damk\"ohler 
        numbers in each case, and we quantify how the coarse and fine tails of 
        the PSD affect the overall kinetics.

  \item We embed the SCM into a dual-porosity hydrological framework of the 
        type used for sedimentary and strongly stratified ores 
        \citep{Robertson2017,Miao2017,Miao2021,Blackmore2018}, and identify 
        the additional dimensionless parameters governing exchange between 
        mobile (advective) and immobile (stagnant or slowly flowing) liquid 
        domains. We then illustrate the combined framework with two simple 
        numerical examples: (i) mapping a lognormal PSD into external- and 
        diffusion-controlled Damk\"ohler distributions, and (ii) an illustrative 
        column/heap example comparing heap-averaged conversion for fine and 
        coarse PSDs under different rate-controlling mechanisms.
\end{enumerate}

The resulting dimensionless formulation is not a replacement for detailed 
numerical models, but a complement. It provides a transparent language for 
interpreting laboratory testwork, for comparing heaps with different PSDs and 
internal pore structures, and for identifying which dimensionless groups must 
be matched to achieve physically meaningful scale-up in hydrometallurgical 
heap leaching.

\section{Analysis of the problem}

\subsection{Geometric similarity, RTD and Damk\"ohler numbers}

Non-ideal reactors, including packed beds and heaps, are often described using residence time 
distributions $E(t)$ and corresponding dimensionless functions 
\citep{Levenspiel1999,Fogler2016,HeapHydroDixon,RTDDrip2023}. For a reactor of volume $V$ and volumetric flow $Q$, 
the mean residence time is $\tau = V/Q$. The dimensionless time and RTD are
\begin{equation}
  \theta = \frac{t}{\tau}, \qquad E^\ast(\theta) = \tau E(t).
\end{equation}
If two reactors are geometrically similar (all length ratios are equal) and dynamically similar 
(e.g.\ same Reynolds, Froude and other relevant dimensionless numbers for the carrier fluid), then the 
dimensionless RTD $E^\ast(\theta)$ of the fluid phase is identical in both units.

Chemical reactions introduce further dimensionless groups, notably Damk\"ohler numbers
\begin{equation}
  \mathrm{Da} = \frac{\text{reaction timescale}}{\text{flow timescale}},
\end{equation}
and, in distributed systems, Peclet numbers comparing advection and dispersion. In heterogeneous systems, 
additional groups characterise intraparticle diffusion (e.g.\ Thiele moduli) and the relative importance 
of mass transfer resistances \citep{Fogler2016,Szekely1975,FoglerSCM,MorenoPulido2025}.

\subsection{Heap leaching as a porous reactor}

Consider a heap of height $H$, irrigated from the top with superficial liquid flux $q$ and porosity 
$\varepsilon$. The interstitial velocity is $u = q/\varepsilon$, and a natural macroscopic time scale 
is the advective residence time
\begin{equation}
  \tau_{\mathrm{heap}} = \frac{H}{u} = \frac{\varepsilon H}{q}.
\end{equation}
We introduce dimensionless variables
\begin{equation}
  Z = \frac{z}{H}, \qquad \Theta = \frac{t}{\tau_{\mathrm{heap}}},
\end{equation}
and write a standard 1D advection--dispersion equation for the reagent concentration $C(z,t)$,
\begin{equation}
  \varepsilon \frac{\partial C}{\partial t}
  + (1-\varepsilon)\,\rho_s \frac{\partial \bar{X}}{\partial t}
  + u \frac{\partial C}{\partial z}
  = \frac{\partial}{\partial z}\left(D_{\mathrm{ax}}\frac{\partial C}{\partial z}\right),
  \label{eq:heap-macro-balance}
\end{equation}
where $\bar{X}(z,t)$ is the local, particle-averaged solid conversion, $D_{\mathrm{ax}}$ is an effective 
axial dispersion coefficient, and $\rho_s$ is the solid density. The corresponding Peclet number is
\begin{equation}
  \mathrm{Pe}_{\mathrm{heap}} = \frac{u H}{D_{\mathrm{ax}}}.
\end{equation}
For two geometrically similar heaps operated at the same $\mathrm{Pe}_{\mathrm{heap}}$ and similar 
saturation/wetting conditions, the dimensionless RTD of the liquid phase is expected to be similar 
\citep{HeapHydroDixon,PreferentialFlow2016,HeapScaleup}.

In addition to the heap-scale length $H$, the system includes a second length scale, the particle diameter 
$d_p$. A fundamental geometric ratio is
\begin{equation}
  \delta = \frac{d_p}{H},
\end{equation}
and the heap is characterised by a PSD $f(d_p)$, often expressed as a 
volume- or mass-based probability density.

If two heaps are geometrically similar at the macroscopic level, but the PSDs differ, then microscopic 
similarity is broken. Even if $\mathrm{Pe}_{\mathrm{heap}}$ and other heap-scale hydrodynamic groups are 
matched, the intraparticle transport and reaction behaviour will differ, and so will the overall 
conversion.

\section{Material and methods}

\subsection{Shrinking-core model for individual particles}

We briefly recall the SCM for spherical particles, which provides a natural 
microscopic description for leaching and gas--solid reactions 
\citep{Szekely1975,FoglerSCM,Fuerstenau2005,MorenoPulido2025}.

Consider a spherical particle of initial radius $R_0 = d_p/2$, density $\rho_s$, and reactive solid A, 
contacted by a reagent in the surrounding fluid at concentration $C_\infty$. In the classical SCM, an 
unreacted core of radius $R(t)$ shrinks with time, surrounded by a product layer through which the reagent 
must diffuse. Depending on the relative importance of external mass transfer, intraparticle diffusion, 
and intrinsic interfacial reaction, different limiting regimes arise.

In the regime where external mass transfer through a fluid film controls the rate, the characteristic time 
scale for conversion follows
\begin{equation}
  t_{\mathrm{ext}}(d_p) = \frac{\rho_s d_p}{6 k_f C_\infty},
  \label{eq:film-time}
\end{equation}
where $k_f$ is a mass transfer coefficient.

The conversion of a single particle may be expressed as
\begin{equation}
  x_p(d_p,t) = g_{\mathrm{ext}}\!\left(\frac{t}{t_{\mathrm{ext}}(d_p)}, C_\infty\right),
\end{equation}
with a known function $g_{\mathrm{ext}}$ depending on the specific SCM formulation and reaction order.

When diffusion through the product layer is rate-limiting, the characteristic time scale is
\begin{equation}
  t_{\mathrm{diff}}(d_p) = \frac{\rho_s d_p^2}{6 D_{\mathrm{eff}} C_\infty},
  \label{eq:diff-time}
\end{equation}
where $D_{\mathrm{eff}}$ is an effective diffusivity in the product layer. The conversion of a particle is
\begin{equation}
  x_p(d_p,t) = g_{\mathrm{diff}}\!\left(\frac{t}{t_{\mathrm{diff}}(d_p)}, C_\infty\right),
\end{equation}
with $g_{\mathrm{diff}}$ again determined by the SCM expressions, e.g.\ the well-known
diffusion-controlled shrinking-core formula 
$1 - 3(1-x)^{2/3} + 2(1-x) = \text{const}\times t$.

Eqs.~\eqref{eq:film-time} and \eqref{eq:diff-time} exhibit the key scale dependencies:
\begin{equation}
  t_{\mathrm{ext}} \propto d_p, \qquad
  t_{\mathrm{diff}} \propto d_p^2.
\end{equation}
Thus, for a given heap-scale time $\tau_{\mathrm{heap}}$, the ratio
\begin{equation}
  R(d_p) = \frac{t_{\mathrm{diff}}}{t_{\mathrm{ext}}}
  = \frac{K_{\mathrm{ext}}}{K_{\mathrm{diff}}} d_p
\end{equation}
varies with $d_p$, where $K_{\mathrm{ext}}$ and $K_{\mathrm{diff}}$ are constants defined below. 
Small particles tend to be more affected by diffusion control, large particles by film control, and 
intermediate sizes may lie in a mixed regime.

\subsection{Particle-scale Damk\"ohler numbers and their distributions}
\label{sec:DaDistributions}

For a heap with characteristic time $\tau_{\mathrm{heap}}$, it is natural to define particle-scale 
Damk\"ohler numbers for the film- and diffusion-controlled mechanisms:
\begin{align}
  \mathrm{Da}_{\mathrm{ext}}(d_p) &= \frac{\tau_{\mathrm{heap}}}{t_{\mathrm{ext}}(d_p)}
  = \frac{6 k_f C_\infty \tau_{\mathrm{heap}}}{\rho_s}\,\frac{1}{d_p}
  = \frac{K_{\mathrm{ext}}}{d_p}, \label{eq:Da-ext} \\
  \mathrm{Da}_{\mathrm{diff}}(d_p) &= \frac{\tau_{\mathrm{heap}}}{t_{\mathrm{diff}}(d_p)}
  = \frac{6 D_{\mathrm{eff}} C_\infty \tau_{\mathrm{heap}}}{\rho_s}\,\frac{1}{d_p^2}
  = \frac{K_{\mathrm{diff}}}{d_p^2}, \label{eq:Da-diff}
\end{align}
with
\begin{equation}
  K_{\mathrm{ext}} = \frac{6 k_f C_\infty \tau_{\mathrm{heap}}}{\rho_s}, \qquad
  K_{\mathrm{diff}} = \frac{6 D_{\mathrm{eff}} C_\infty \tau_{\mathrm{heap}}}{\rho_s}.
\end{equation}
For a given dimensionless heap time $\Theta = t/\tau_{\mathrm{heap}}$, one can write
\begin{equation}
  x_p(d_p,\Theta) = 
  \begin{cases}
    g_{\mathrm{ext}}\big(\Theta\,\mathrm{Da}_{\mathrm{ext}}(d_p), C_\infty\big),
      & \text{film control}, \\
    g_{\mathrm{diff}}\big(\Theta\,\mathrm{Da}_{\mathrm{diff}}(d_p), C_\infty\big),
      & \text{diffusion control}.
  \end{cases}
\end{equation}
The heap-averaged conversion at time $\Theta$ is then
\begin{equation}
  X_{\mathrm{heap}}(\Theta) = 
  \int_0^\infty x_p(d_p,\Theta)\, f(d_p)\,dd_p,
\end{equation}
where $f(d_p)$ is the PSD.

Let $f(d_p)$ be a normalised PSD,
\begin{equation}
  \int_0^\infty f(d_p)\,dd_p = 1.
\end{equation}
We wish to determine the induced probability density $g(\mathrm{Da})$ of a Damk\"ohler number of interest, 
using change-of-variable formulas.

\subsubsection{External film control}

For external control, $\mathrm{Da} = \mathrm{Da}_{\mathrm{ext}} = K_{\mathrm{ext}}/d_p$, so
\begin{equation}
  d_p = \frac{K_{\mathrm{ext}}}{\mathrm{Da}}, \qquad
  \frac{d d_p}{d\mathrm{Da}} = -\frac{K_{\mathrm{ext}}}{\mathrm{Da}^2}.
\end{equation}
Thus, the pdf $g_{\mathrm{ext}}(\mathrm{Da})$ is
\begin{equation}
  g_{\mathrm{ext}}(\mathrm{Da})
  = f\!\left(\frac{K_{\mathrm{ext}}}{\mathrm{Da}}\right)
    \left|\frac{d d_p}{d\mathrm{Da}}\right|
  = f\!\left(\frac{K_{\mathrm{ext}}}{\mathrm{Da}}\right)
    \frac{K_{\mathrm{ext}}}{\mathrm{Da}^2}.
  \label{eq:g-ext}
\end{equation}
Even if $f(d_p)$ is smooth and unimodal, the factor $\mathrm{Da}^{-2}$ can skew the Da-distribution.

\subsubsection{Intraparticle diffusion control}

For diffusion control, $\mathrm{Da} = \mathrm{Da}_{\mathrm{diff}} = K_{\mathrm{diff}}/d_p^2$, so
\begin{equation}
  d_p = \sqrt{\frac{K_{\mathrm{diff}}}{\mathrm{Da}}}, \qquad
  \frac{d d_p}{d\mathrm{Da}} = -\frac{1}{2}\sqrt{K_{\mathrm{diff}}}\,\mathrm{Da}^{-3/2}.
\end{equation}
The pdf $g_{\mathrm{diff}}(\mathrm{Da})$ becomes
\begin{equation}
  g_{\mathrm{diff}}(\mathrm{Da})
  = f\!\left(\sqrt{\frac{K_{\mathrm{diff}}}{\mathrm{Da}}}\right)
    \frac{1}{2}\sqrt{K_{\mathrm{diff}}}\,\mathrm{Da}^{-3/2}.
  \label{eq:g-diff}
\end{equation}
Here the weighting $\mathrm{Da}^{-3/2}$ and the square-root mapping in the argument of $f$ make the 
resulting Da-distribution even more sensitive to the small-particle (large-Da) tail of the PSD than in 
the film-controlled case.

\subsubsection{Mixed control (series resistances)}

When both external mass transfer and intraparticle diffusion contribute significantly, the SCM becomes 
more complex \citep{Szekely1975,FoglerSCM,MorenoPulido2025}. For scaling and similarity arguments, a 
useful approximation is to treat the two mechanisms as resistances in series, with an effective 
characteristic time satisfying
\begin{equation}
  \frac{1}{t_{\mathrm{char}}} \approx \frac{1}{t_{\mathrm{ext}}} + \frac{1}{t_{\mathrm{diff}}}.
\end{equation}
This leads to an effective Damk\"ohler number
\begin{equation}
  \mathrm{Da}_{\mathrm{eff}}(d_p)
  = \frac{\tau_{\mathrm{heap}}}{t_{\mathrm{char}}(d_p)}
  \approx \mathrm{Da}_{\mathrm{ext}}(d_p) + \mathrm{Da}_{\mathrm{diff}}(d_p)
  = \frac{K_{\mathrm{ext}}}{d_p} + \frac{K_{\mathrm{diff}}}{d_p^2}.
  \label{eq:Da-eff}
\end{equation}

To construct the corresponding distribution $g_{\mathrm{mix}}(\mathrm{Da})$, we solve for $d_p$ as a 
function of $\mathrm{Da}$. Multiplying Eq.~\eqref{eq:Da-eff} by $d_p^2$,
\begin{equation}
  \mathrm{Da}\,d_p^2 = K_{\mathrm{ext}} d_p + K_{\mathrm{diff}},
\end{equation}
so $d_p$ satisfies a quadratic equation
\begin{equation}
  \mathrm{Da}\,d_p^2 - K_{\mathrm{ext}} d_p - K_{\mathrm{diff}} = 0.
\end{equation}
The physically relevant (positive) root is
\begin{equation}
  d_p(\mathrm{Da})
  = \frac{K_{\mathrm{ext}} + \sqrt{K_{\mathrm{ext}}^2 + 4 K_{\mathrm{diff}} \mathrm{Da}}}{2\,\mathrm{Da}}.
  \label{eq:dp-mix}
\end{equation}
Differentiating $\mathrm{Da}_{\mathrm{eff}}(d_p)$ with respect to $d_p$ gives
\begin{equation}
  \frac{d\,\mathrm{Da}}{dd_p}
  = -\frac{K_{\mathrm{ext}}}{d_p^2} - \frac{2K_{\mathrm{diff}}}{d_p^3}
  = -\frac{K_{\mathrm{ext}} d_p + 2K_{\mathrm{diff}}}{d_p^3},
\end{equation}
and hence
\begin{equation}
  \left|\frac{dd_p}{d\mathrm{Da}}\right|
  = \frac{d_p^3}{K_{\mathrm{ext}} d_p + 2K_{\mathrm{diff}}}.
\end{equation}
The mixed-control Damk\"ohler pdf is therefore
\begin{equation}
  g_{\mathrm{mix}}(\mathrm{Da})
  = f\big(d_p(\mathrm{Da})\big)\,
    \frac{d_p(\mathrm{Da})^3}{K_{\mathrm{ext}} d_p(\mathrm{Da}) + 2K_{\mathrm{diff}}},
  \label{eq:g-mix}
\end{equation}
with $d_p(\mathrm{Da})$ given by Eq.~\eqref{eq:dp-mix}.

\subsection{Coupling shrinking-core kinetics to dual-porosity flow}

Many ore bodies, particularly sedimentary or strongly stratified deposits, exhibit internal 
structure that naturally suggests a dual-porosity or dual-pore-system description: relatively 
open macropores or preferential channels that carry most of the advective flow, and a larger 
volume of microporous matrix where most of the mineral resides and flow is slow, diffusive, 
or quasi-stagnant. This picture is closely related to dual-porosity and dual-permeability 
models developed in petroleum and hydrogeology 
\citep{Barenblatt1960,WarrenRoot1963,Zimmerman1993,DoublePorosity2020}, and has been adapted to heap 
leaching in 1D column models and 3D pore-scale simulations 
\citep{Robertson2017,Miao2017,Miao2021,Blackmore2018,HeapScaleup}.

In a dual-porosity interpretation of heap leaching, the void space is partitioned into:
\begin{itemize}
  \item a mobile (advective) domain with porosity $\varepsilon_m$ and concentration $C_m(z,t)$, and
  \item an immobile (stagnant or slowly moving) domain with porosity $\varepsilon_{im}$ and concentration 
        $C_{im}(z,t)$,
\end{itemize}
with $\varepsilon_m + \varepsilon_{im} \approx \varepsilon$. The solid is mostly associated with the immobile 
domain, where diffusion-controlled SCM kinetics are often most relevant.

A widely used structure for dual-porosity models in single-phase flow is to write coupled balances for 
$C_m$ and $C_{im}$ with an interporosity mass-transfer term of the form 
$\alpha (C_m - C_{im})$ \citep{Barenblatt1960,WarrenRoot1963,Zimmerman1993,Robertson2017}. For a 1D heap, 
one may write
\begin{align}
  \varepsilon_m \frac{\partial C_m}{\partial t}
  + u \frac{\partial C_m}{\partial z}
  &= \frac{\partial}{\partial z}\!\left(D_{\mathrm{ax},m} \frac{\partial C_m}{\partial z}\right)
     - \alpha (C_m - C_{im}), \label{eq:dual-mob} \\
  \varepsilon_{im} \frac{\partial C_{im}}{\partial t}
  + (1-\varepsilon)\,\rho_s \frac{\partial \bar{X}}{\partial t}
  &= \alpha (C_m - C_{im}), \label{eq:dual-imm}
\end{align}
where $D_{\mathrm{ax},m}$ is an axial dispersion coefficient in the mobile domain, and 
$\alpha$ is an interporosity exchange coefficient (units of s$^{-1}$).

Introducing the macroscopic time and space scales,
\begin{equation}
  Z = \frac{z}{H}, \qquad \Theta = \frac{t}{\tau_{\mathrm{heap}}}, \qquad
  \tau_{\mathrm{heap}} = \frac{H}{u},
\end{equation}
and defining dimensionless concentrations $\hat{C}_m=C_m/C_\infty$, $\hat{C}_{im}=C_{im}/C_\infty$, 
the mobile-domain balance becomes
\begin{equation}
  \varepsilon_m \frac{\partial \hat{C}_m}{\partial \Theta}
  + \frac{\partial \hat{C}_m}{\partial Z}
  = \frac{1}{\mathrm{Pe}_m}\frac{\partial^2 \hat{C}_m}{\partial Z^2}
    - \Lambda_{\mathrm{ex}} \left(\hat{C}_m - \hat{C}_{im}\right),
  \label{eq:dual-mob-dim}
\end{equation}
with
\begin{equation}
  \mathrm{Pe}_m = \frac{u H}{D_{\mathrm{ax},m}}, \qquad
  \Lambda_{\mathrm{ex}} = \alpha \tau_{\mathrm{heap}}.
\end{equation}
Similarly, the immobile-domain balance becomes
\begin{equation}
  \varepsilon_{im} \frac{\partial \hat{C}_{im}}{\partial \Theta}
  + (1-\varepsilon)\,\rho_s \frac{C_\infty^{-1}}{\tau_{\mathrm{heap}}} \frac{\partial \bar{X}}{\partial \Theta}
  = \Lambda_{\mathrm{ex}} \left(\hat{C}_m - \hat{C}_{im}\right).
  \label{eq:dual-imm-dim}
\end{equation}
It is common to introduce a capacity ratio
\begin{equation}
  \omega = \frac{\varepsilon_{im}}{\varepsilon_m},
\end{equation}
and a dimensionless exchange Damk\"ohler number
\begin{equation}
  \mathrm{Da}_{\mathrm{ex}} = \Lambda_{\mathrm{ex}} = \alpha \tau_{\mathrm{heap}}.
\end{equation}
With these definitions, Eqs.~\eqref{eq:dual-mob-dim}--\eqref{eq:dual-imm-dim} show that a dual-porosity heap 
is characterised, at the macroscopic level, by $\mathrm{Pe}_m$, $\omega$ and $\mathrm{Da}_{\mathrm{ex}}$, 
in addition to the particle-scale Damk\"ohler numbers of Sec.~\ref{sec:DaDistributions}.

\subsection{Parameter-estimation workflow (illustrative example)}
\label{sec:estimation}

Although this work is primarily a similarity and scaling analysis, the same
dimensionless structure supports practical parameter estimation from column
tests. The objective is to infer a minimal set of \emph{mechanistic} parameters
(e.g.\ $k_f$, $D_{\mathrm{eff}}$, $\alpha$, $D_{\mathrm{ax}}$) or, equivalently,
their dimensionless combinations (e.g.\ $K_{\mathrm{ext}}$, $K_{\mathrm{diff}}$,
$\mathrm{Da}_{\mathrm{ex}}$, $\mathrm{Pe}_m$), from tracer RTD experiments and
leaching conversion data.

\subsubsection{Step A: Hydrodynamic calibration from RTD}
A conservative tracer test provides the outlet RTD $E(t)$, from which the mean
residence time $\tau$ and the dimensionless RTD $E^\ast(\theta)=\tau E(t)$ are
computed \citep{Levenspiel1999}. A 1D advection--dispersion (or dual-porosity)
model is then calibrated to the tracer response to estimate hydraulic
parameters:
\begin{itemize}
  \item Single-domain: estimate $D_{\mathrm{ax}}$ (thus $\mathrm{Pe}_{\mathrm{heap}}=uH/D_{\mathrm{ax}}$).
  \item Dual-porosity: estimate $(D_{\mathrm{ax},m},\alpha,\omega)$, thus
        $(\mathrm{Pe}_m,\mathrm{Da}_{\mathrm{ex}},\omega)$.
\end{itemize}
This separation is useful because tracer RTD fitting is largely independent of
leaching kinetics, improving identifiability.

\subsubsection{Step B: Particle-scale kinetic calibration from leaching curves}
Let $X_{\mathrm{heap}}(t)$ denote the measured heap-averaged solid conversion in
a column (or a representative segment of a heap). Given a measured PSD $f(d_p)$
and calibrated hydrodynamics from Step A, kinetic parameters can be inferred by
minimising a least-squares objective on \emph{dimensionless} time:
\begin{equation}
  \min_{\boldsymbol{p}}\;
  \sum_{i=1}^{N}
  \left[
    X_{\mathrm{heap}}^{\mathrm{model}}(\Theta_i;\boldsymbol{p})
    - X_{\mathrm{heap}}^{\mathrm{data}}(\Theta_i)
  \right]^2,
  \qquad \Theta_i = t_i/\tau_{\mathrm{heap}},
\end{equation}
where $\boldsymbol{p}$ may be chosen as $(k_f,D_{\mathrm{eff}})$ or equivalently
$(K_{\mathrm{ext}},K_{\mathrm{diff}})$. Under external film control, the model
depends primarily on $K_{\mathrm{ext}}$ through $\mathrm{Da}_{\mathrm{ext}}(d_p)$
[Eq.~\eqref{eq:Da-ext}], whereas under diffusion control it depends primarily on
$K_{\mathrm{diff}}$ through $\mathrm{Da}_{\mathrm{diff}}(d_p)$
[Eq.~\eqref{eq:Da-diff}]. In mixed control, both parameters influence the fit
and can be partially correlated; the distributional mapping
[Eqs.~\eqref{eq:g-ext}, \eqref{eq:g-diff}, \eqref{eq:g-mix}] clarifies which PSD
fractions constrain which resistance.

\subsubsection{Step C: Consistency checks and uncertainty}
A practical diagnostic is to repeat the fit under (i) film-only, (ii)
diffusion-only, and (iii) mixed control, and compare residual structure. For
example, a persistent late-time underprediction under film-only kinetics often
indicates diffusion or immobile-domain limitations, whereas early-time
underprediction under diffusion-only kinetics may indicate finite external mass
transfer. Uncertainty can be quantified by bootstrap resampling of the time
points or by estimating the local curvature (Hessian) of the objective near the
optimum, reporting confidence intervals on $(K_{\mathrm{ext}},K_{\mathrm{diff}})$
and, if dual porosity is used, on $(\omega,\mathrm{Da}_{\mathrm{ex}})$.

\subsubsection{Workflow summary}
In compact form, a recommended workflow is:
\begin{center}
\begin{tabular}{l}
\hline
(1) Measure PSD $f(d_p)$ and column geometry/irrigation to define $\tau_{\mathrm{heap}}$. \\
(2) Run tracer RTD $\Rightarrow$ fit hydrodynamics ($\mathrm{Pe}$, and if needed $\omega,\mathrm{Da}_{\mathrm{ex}}$). \\
(3) Run leaching $X_{\mathrm{heap}}(t)$ $\Rightarrow$ fit $(K_{\mathrm{ext}},K_{\mathrm{diff}})$ (or $k_f,D_{\mathrm{eff}}$). \\
(4) Validate by predicting a second irrigation condition or a second PSD. \\
\hline
\end{tabular}
\end{center}

\section{Results and discussion}

\subsection{Numerical illustrations}

To illustrate the preceding framework, we present two simple numerical examples. 
The goal is not to fit any particular experimental system, but rather to 
demonstrate how the PSD and the choice of rate-controlling mechanism influence 
the distributions of Damk\"ohler numbers and the heap-averaged conversion in a 
dimensionless setting.

\subsubsection{Example 1: lognormal PSD and Damk\"ohler distributions}

We consider a lognormal PSD for the particle diameter,
\begin{equation}
  f(d_p) = \frac{1}{d_p \sigma_{\ln}\sqrt{2\pi}}
           \exp\!\left[-\frac{\big(\ln d_p - \mu_{\ln}\big)^2}{2\sigma_{\ln}^2}\right],
\end{equation}
with median $d_{50} = 10~\text{mm}$ and geometric standard deviation 
$\sigma_g=\exp(\sigma_{\ln})\approx 1.65$ ($\sigma_{\ln}=0.5$). We discretise 
$d_p\in[1,50]$~mm on a fine grid and normalise $f(d_p)$ numerically.

We choose dimensionless constants $K_{\mathrm{ext}}=10^{-4}$ and 
$K_{\mathrm{diff}}=10^{-6}$. For the chosen PSD, the particle-scale Damk\"ohler 
numbers
\begin{equation}
  \mathrm{Da}_{\mathrm{ext}}(d_p)=\frac{K_{\mathrm{ext}}}{d_p}, 
  \qquad
  \mathrm{Da}_{\mathrm{diff}}(d_p)=\frac{K_{\mathrm{diff}}}{d_p^2}
\end{equation}
are of order $10^{-3}$--$10^{-1}$ over the size range of interest, corresponding 
to moderately slow reaction relative to the heap residence time. Using the 
change-of-variable formulas in Eqs.~\eqref{eq:g-ext} and \eqref{eq:g-diff}, we 
construct the induced pdfs $g_{\mathrm{ext}}(\mathrm{Da})$ and 
$g_{\mathrm{diff}}(\mathrm{Da})$ by mapping the $d_p$ grid to $\mathrm{Da}$ 
and renormalising.

For these parameter values, we obtain approximate statistics
\begin{align*}
  \langle \mathrm{Da}_{\mathrm{ext}} \rangle &\approx 3.1\times 10^{-2}, &
  \mathrm{Da}_{\mathrm{ext},5\%} &\approx 4.4\times 10^{-3}, &
  \mathrm{Da}_{\mathrm{ext},95\%} &\approx 2.3\times 10^{-2},\\
  \langle \mathrm{Da}_{\mathrm{diff}} \rangle &\approx 4.5\times 10^{-1}, &
  \mathrm{Da}_{\mathrm{diff},5\%} &\approx 1.9\times 10^{-3}, &
  \mathrm{Da}_{\mathrm{diff},95\%} &\approx 5.2\times 10^{-2}.
\end{align*}

The diffusion-controlled distribution is much more skewed than the film-controlled one: 
although 90\% of $\mathrm{Da}_{\mathrm{diff}}$ values lie between 
$\approx 2\times 10^{-3}$ and $\approx 5\times 10^{-2}$, the mean is an order of magnitude 
larger, reflecting a long upper tail dominated by the smallest particles.

\begin{figure}[H]
  \centering
  \includegraphics[width=0.78\textwidth]{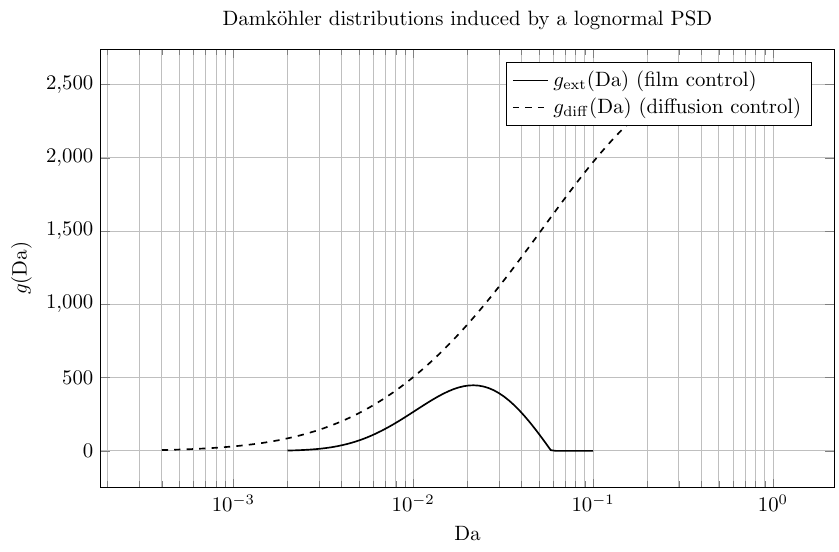}
  \caption{Damk\"ohler distributions $g_{\mathrm{ext}}(\mathrm{Da})$ and
  $g_{\mathrm{diff}}(\mathrm{Da})$ induced by a lognormal PSD with $d_{50}=10$~mm
  and $\sigma_{\ln}=0.5$, for $K_{\mathrm{ext}}=10^{-4}$ and
  $K_{\mathrm{diff}}=10^{-6}$. The diffusion-controlled case  reveals a broader,
  more skewed distribution with a heavier upper (high-Da) tail.}
  \label{fig:gDa}
\end{figure}

Figure~\ref{fig:gDa} compares $g_{\mathrm{ext}}(\mathrm{Da})$ and 
$g_{\mathrm{diff}}(\mathrm{Da})$ on semi-logarithmic axes. The film-controlled 
distribution is relatively narrow, whereas the diffusion-controlled distribution exhibits 
a pronounced right tail. This quantitatively illustrates that, for a given PSD, 
diffusion-controlled kinetics are more sensitive to the presence of a fine fraction, which 
can dominate the overall reaction rate despite contributing a small fraction of the total 
solid mass.

This behaviour follows directly from the mapping $\mathrm{Da}_{\mathrm{diff}}\propto d_p^{-2}$:
even a modest fine fraction translates into a disproportionately large
high-$\mathrm{Da}$ population. As a consequence, diffusion-controlled heaps can
appear kinetically ``fast'' at early times while simultaneously remaining
sensitive to the slow leaching of coarse particles at late times, motivating
explicit reporting (and, when feasible, preservation) of PSD tails in scale-up studies.

\subsubsection{Example 2: illustrative heap conversions for fine and coarse PSDs}

As a second illustration, consider two lognormal PSDs with the same geometric standard deviation 
$\sigma_{\ln}=0.5$ but different medians: a ``fine'' PSD with $d_{50}=8$~mm, and a ``coarse'' PSD 
with $d_{50}=20$~mm. We again take $K_{\mathrm{ext}}=10^{-4}$ and $K_{\mathrm{diff}}=10^{-6}$, 
and discretise $d_p\in[1,50]$~mm with the same numerical resolution.

For each PSD, we compute the corresponding Damk\"ohler numbers 
$\mathrm{Da}_{\mathrm{ext}}(d_p)$ and $\mathrm{Da}_{\mathrm{diff}}(d_p)$ and then approximate 
the heap-averaged conversion at dimensionless time $\Theta$ using simple surrogate SCM-like 
expressions:
\begin{align}
  X_{\mathrm{heap}}^{(\mathrm{film})}(\Theta)
  &\approx \int_0^\infty \left[1-\exp\big(-\Theta\,\mathrm{Da}_{\mathrm{ext}}(d_p)\big)\right]
   f(d_p)\,dd_p, \label{eq:X-film-example}\\
  X_{\mathrm{heap}}^{(\mathrm{diff})}(\Theta)
  &\approx \int_0^\infty \left[1-\exp\big(-\Theta\,\sqrt{\mathrm{Da}_{\mathrm{diff}}(d_p)}\big)\right]
   f(d_p)\,dd_p. \label{eq:X-diff-example}
\end{align}
These surrogate forms mimic the qualitative behaviour of SCM kinetics in film- and diffusion-controlled 
regimes while keeping the numerical example analytically simple.

\begin{figure}[H]
  \centering
  \includegraphics[width=0.78\textwidth]{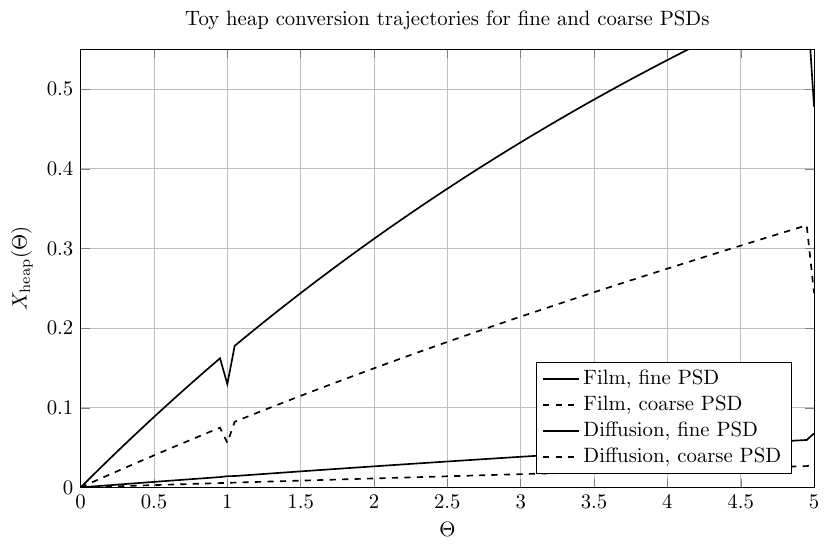}
  \caption{Heap-averaged conversion trajectories $X_{\mathrm{heap}}(\Theta)$ for
  fine ($d_{50}=8$~mm) and coarse ($d_{50}=20$~mm) PSDs under film- and
  diffusion-controlled surrogate kinetics [Eqs.~\eqref{eq:X-film-example}--\eqref{eq:X-diff-example}].
  The diffusion-controlled cases show both higher conversions and a stronger dependence on PSD.}
  \label{fig:Xheap}
\end{figure}

Figure~\ref{fig:Xheap} shows the resulting trajectories $X_{\mathrm{heap}}(\Theta)$ for 
$\Theta\in[0,5]$. At representative times $\Theta=1$ and $\Theta=5$, we obtain the approximate values
\begin{center}
\begin{tabular}{lcc}
\hline
& $X_{\mathrm{heap}}(\Theta=1)$ & $X_{\mathrm{heap}}(\Theta=5)$ \\
\hline
Film control, fine PSD   & $0.014$ & $0.068$ \\
Film control, coarse PSD & $0.0058$ & $0.028$ \\
Diffusion control, fine PSD   & $0.13$ & $0.48$ \\
Diffusion control, coarse PSD & $0.056$ & $0.24$ \\
\hline
\end{tabular}
\end{center}

Several trends emerge:
\begin{itemize}
  \item For both mechanisms, the fine PSD leaches faster than the coarse PSD, as expected from 
        Eqs.~\eqref{eq:Da-ext}--\eqref{eq:Da-diff}.
  \item The diffusion-controlled cases show significantly higher conversions at fixed $\Theta$ than the 
        film-controlled cases for the chosen parameters, reflecting the larger typical 
        $\mathrm{Da}_{\mathrm{diff}}$ values.
  \item The \emph{relative} penalty for coarsening the PSD is more pronounced when diffusion dominates: 
        at $\Theta=5$, the coarse-PSD conversion is about 41\% of the fine-PSD value under film control 
        and about 51\% under diffusion control, but the absolute gap in $X$ is much larger in the latter. 
        This is consistent with the stronger $d_p$-dependence of $\mathrm{Da}_{\mathrm{diff}}\propto d_p^{-2}$.
\end{itemize}

A practical implication is that a single ``representative'' particle size is
more defensible under film control than under diffusion control. When diffusion
dominates, the PSD induces a broad distribution of characteristic times, so
late-time recovery is governed by the coarse tail while early-time response can
be dominated by fines. This separation of time scales is further amplified if
dual porosity limits reagent access to the reactive matrix.

\subsection{Similarity criteria and sensitivity to PSD and dual porosity}

The expressions above allow us to state precise similarity conditions for two geometrically 
similar heaps, A and B, that differ in PSD and internal pore structure. Suppose that:
\begin{itemize}
  \item The heaps have the same macroscopic geometry (same aspect ratios, slope angles, etc.).
  \item They are operated with the same fluid properties, reagent concentration $C_\infty$, and 
        temperature.
  \item The macroscopic hydrodynamics are dynamically similar (same $\mathrm{Pe}_{\mathrm{heap}}$ or 
        $\mathrm{Pe}_m$, similar saturation and capillary numbers).
  \item The dual-porosity structure is similar in the dimensionless sense ($\omega$ and 
        $\mathrm{Da}_{\mathrm{ex}}$ matched).
\end{itemize}
Then the dimensionless RTD of the liquid phase and the partitioning between mobile and immobile 
domains are the same in both heaps. However, unless their \emph{dimensionless} PSDs match 
(e.g.\ the PSD scaled by heap height, $d_p/H$, is identical), the distributions of particle-scale 
Damk\"ohler numbers $g_{\mathrm{ext}}(\mathrm{Da})$, $g_{\mathrm{diff}}(\mathrm{Da})$, or 
$g_{\mathrm{mix}}(\mathrm{Da})$, and their ratios to $\mathrm{Da}_{\mathrm{ex}}$, will in general differ.

Because $\mathrm{Da}_{\mathrm{diff}} \propto 1/d_p^2$, diffusion-controlled heaps are significantly more 
sensitive to changes in PSD tails than film-controlled heaps, where 
$\mathrm{Da}_{\mathrm{ext}} \propto 1/d_p$. In mixed-control systems, the relative importance of film and 
diffusion contributions varies across the PSD, which can lead to complex responses of heap performance 
to changes in crushing or agglomeration practices. Dual-porosity structure further modulates this picture 
by controlling how effectively reagent in the mobile domain can access the immobile, reactive matrix.

From a scale-up perspective, this analysis suggests that matching only macroscopic hydrodynamic 
dimensionless groups is generally insufficient. For true similarity in heap leaching performance, one 
must also match the PSD in dimensionless form and, if possible, the distributions of relevant 
Damk\"ohler numbers implied by the SCM, as well as the dual-porosity parameters $\omega$ and 
$\mathrm{Da}_{\mathrm{ex}}$. This observation is consistent with more 
detailed, numerically intensive models of heap leaching that resolve both hydrology and intraparticle 
transport \citep{DixonPetersen2003,DixonPetersen2004,Robertson2017,Miao2017,Miao2021,HeapScaleup,HeapSim2D}.

\begin{table}[H]
  \centering
  \small
  \caption{Summary of key dimensionless groups used in the analysis.}
  \label{tab:dimless-groups}
  \begin{tabular}{lll}
    \hline
    Symbol & Definition & Meaning \\
    \hline
    $Z$ & $z/H$ & Dimensionless vertical coordinate. \\
    $\Theta$ & $t/\tau_{\mathrm{heap}}$ & Dimensionless time. \\
    $\delta$ & $d_p/H$ & Particle-to-heap size ratio. \\
    $\varepsilon$ & -- & Total heap porosity. \\
    $\varepsilon_m$ & -- & Mobile-domain porosity. \\
    $\varepsilon_{im}$ & -- & Immobile-domain porosity. \\
    $\omega$ & $\varepsilon_{im}/\varepsilon_m$ & Immobile/mobile capacity ratio. \\
    $\mathrm{Pe}_{\mathrm{heap}}$ & $uH/D_{\mathrm{ax}}$ & Heap Peclet number. \\
    $\mathrm{Pe}_m$ & $uH/D_{\mathrm{ax},m}$ & Mobile-domain Peclet number. \\
    $\mathrm{Da}_{\mathrm{ext}}(d_p)$ & $K_{\mathrm{ext}}/d_p$ & Film-control Damk\"ohler ($\propto d_p^{-1}$). \\
    $\mathrm{Da}_{\mathrm{diff}}(d_p)$ & $K_{\mathrm{diff}}/d_p^2$ & Diffusion-control Damk\"ohler ($\propto d_p^{-2}$). \\
    $\mathrm{Da}_{\mathrm{eff}}(d_p)$ & $\approx \mathrm{Da}_{\mathrm{ext}}+\mathrm{Da}_{\mathrm{diff}}$ & Effective Damk\"ohler (mixed control). \\
    $\mathrm{Da}_{\mathrm{ex}}$ & $\alpha \tau_{\mathrm{heap}}$ & Mobile--immobile exchange Damk\"ohler. \\
    \hline
  \end{tabular}
\end{table}

\section{Conclusions}

We developed a dimensionless framework that separates hydrodynamic similarity
(from RTD/transport) from microscopic similarity (from PSD-dependent particle-scale kinetics),
and shows how both must be matched for robust heap-leach scale-up. While the derivations draw on standard concepts from 
chemical reaction engineering, the emphasis throughout has been on quantities 
that are directly relevant to column testing and industrial heap design.

At the macroscopic level, we highlighted how geometric and dynamic similarity 
in the hydrology (e.g.\ matching heap geometry, irrigation conditions, and 
Peclet and capillary numbers) leads to similar dimensionless RTDs in 
geometrically similar heaps. For heaps that exhibit dual-porosity behaviour, 
we introduced additional dimensionless parameters---a capacity ratio between 
mobile and immobile liquid domains and an interporosity Damk\"ohler number 
$\mathrm{Da}_{\mathrm{ex}}$---which together control how rapidly reagent in 
advective channels can access the reactive matrix. These parameters naturally 
complement the dimensionless groups used in more detailed dual-porosity heap 
models reported in the hydrometallurgical literature 
\citep{Robertson2017,Miao2017,Miao2021,Blackmore2018,HeapScaleup}.

At the microscopic level, we used the shrinking-core model to derive explicit 
expressions for how a PSD in particle diameter maps into distributions of 
Damk\"ohler numbers under external film control, intraparticle diffusion 
control, and mixed control. These formulas make quantitatively clear that:
\begin{itemize}
  \item For film-controlled systems, the relevant Damk\"ohler number scales as 
        $d_p^{-1}$, so coarsening the PSD penalises leaching rates, but the 
        sensitivity is moderate.
  \item For diffusion-controlled systems, the Damk\"ohler number scales as 
        $d_p^{-2}$, leading to a much stronger dependence on PSD tails and a 
        heavy-tailed distribution of particle-scale Damk\"ohler numbers.
  \item In mixed-control systems, each particle is characterised by a pair of 
        Damk\"ohler numbers (for film and diffusion) or an effective 
        Damk\"ohler number and a resistance ratio. The PSD therefore induces a 
        one-dimensional curve in this two-dimensional parameter space, and the 
        impact of changing the PSD depends on how this curve spans regions of 
        film-, diffusion- and mixed-control behaviour.
\end{itemize}

Simple numerical examples were used to illustrate these points. For a 
lognormal PSD with parameters typical of crushed ore, the diffusion-controlled 
Damk\"ohler distribution was found to be considerably more skewed than the 
film-controlled distribution, with the smallest particles exerting a 
disproportionate influence on the mean Damk\"ohler number. An illustrative heap example 
showed how fine and coarse PSDs produce markedly different heap-averaged 
conversion trajectories under both mechanisms, with diffusion-controlled 
systems exhibiting higher conversions but also larger absolute penalties when 
the PSD is coarsened.

From a practical standpoint, the analysis suggests several implications for 
hydrometallurgical heap leaching:
\begin{itemize}
  \item Matching only macroscopic hydraulic conditions (e.g.\ irrigation rate, 
        solution chemistry, overall RTD) between column tests and full-scale 
        heaps is not sufficient for predictive scale-up if the PSD and 
        dual-porosity structure differ significantly.
  \item Laboratory testwork intended to support design of industrial heaps 
        should, where possible, preserve not only the PSD but also the key 
        dual-porosity parameters (mobile/immobile capacity ratio, interporosity 
        exchange rate) and operating regimes that determine whether film, 
        diffusion or mixed control is dominant.
  \item The dimensionless groups identified here provide a compact way to 
        document and compare column and heap conditions, to assess whether two 
        systems are ``similar enough'' for direct transfer of kinetic 
        information, and to design sensitivity studies on PSD and ore type.
\end{itemize}

The present framework could be extended in several directions of interest for 
future work: coupling to microbially mediated kinetics in bioleaching, 
incorporating more general intraparticle reaction--diffusion models or 
multi-mineral ore textures, and embedding the dimensionless structure into 
existing numerical simulators such as HeapSim or dual-porosity flow codes for 
parameter screening and model reduction. Ultimately, we hope that expressing 
heap leaching in terms of a small set of well-defined dimensionless groups 
will aid practitioners in designing more informative testwork, in understanding 
the impact of PSD and ore variability, and in achieving more robust and 
transparent scale-up in hydrometallurgical heap operations.

\section*{Nomenclature (dimensional parameters)}

\noindent\textbf{Geometry and hydrodynamics}
\begin{center}
\small
\begin{tabular}{ll}
$H$ & heap (or column) height [m] \\
$z$ & vertical coordinate [m] \\
$V$ & reactor/heap control volume [m$^3$] \\
$q$ & superficial liquid flux [m s$^{-1}$] \\
$u$ & interstitial velocity, $u=q/\varepsilon$ [m s$^{-1}$] \\
$\tau_{\mathrm{heap}}$ & advective time scale, $\varepsilon H/q$ [s] \\
$D_{\mathrm{ax}}$ & effective axial dispersion coefficient [m$^2$ s$^{-1}$] \\
$D_{\mathrm{ax},m}$ & axial dispersion in mobile domain [m$^2$ s$^{-1}$] \\
\end{tabular}
\end{center}

\noindent\textbf{Solid and particle-scale transport}
\begin{center}
\small
\begin{tabular}{ll}
$d_p$ & particle diameter [m] \\
$R_0$ & initial particle radius ($d_p/2$) [m] \\
$\rho_s$ & solid density [kg m$^{-3}$] \\
$D_{\mathrm{eff}}$ & effective diffusivity in product layer [m$^2$ s$^{-1}$] \\
$k_f$ & external mass-transfer coefficient [m s$^{-1}$] \\
\end{tabular}
\end{center}

\noindent\textbf{Chemistry and dual porosity}
\begin{center}
\small
\begin{tabular}{ll}
$C_\infty$ & bulk reagent concentration [mol m$^{-3}$] \\
$C_m,\,C_{im}$ & mobile and immobile concentrations [mol m$^{-3}$] \\
$\alpha$ & mobile--immobile exchange coefficient [s$^{-1}$] \\
$\varepsilon$ & total porosity [-] \\
$\varepsilon_m,\,\varepsilon_{im}$ & mobile/immobile porosities [-] \\
\end{tabular}
\end{center}

\section*{Acknowledgements}

[Text omitted.]

\bibliographystyle{plainnat}

\end{document}